\documentclass[english]{article}
\usepackage[T1]{fontenc}
\usepackage[latin9]{inputenc}
\usepackage{color}
\usepackage{amssymb}
\usepackage{graphicx}
\usepackage{esint}

\makeatletter

\newcommand*\LyXThinSpace{\,\hspace{0pt}}

\newcommand{\lyxaddress}[1]{
	\par {\raggedright #1
	\vspace{1.4em}
	\noindent\par}
}

\makeatother

\usepackage{babel}
\begin{document}
\title{Quantum Anomalous Semimetals}
\author{Bo Fu, Jin-Yu Zou, Zi-Ang Hu, Huan-Wen Wang, and Shun-Qing Shen{*}}
\maketitle

\lyxaddress{Department of Physics, The University of Hong Kong, Pokfulam Road, Hong Kong, China}

\paragraph*{Summary}

The topological states of matter and topological materials have been attracting extensive
interests as one of the frontier topics in condensed matter physics and materials
science since the discovery of quantum Hall effect in 1980s. So far all the topological
phases such as quantum Hall effect, quantum spin Hall effect and topological insulators
and superconductors are characterized by a nonzero integer or $Z$ and $Z_{2}$ topological
invariant. None is a half-integer or fractional. Here we propose a novel type of
semimetals which hosts a single cone of Wilson fermions instead of Dirac fermions.
The Wilson fermions possess linear dispersion near the energy crossing point, but
breaks the chiral or parity symmetry such that an unpaired Dirac cone can be realized
on a lattice. They are not prohibited by the Nielsen-Ninomiya theorem and avoid the
fermion doubling problem. We find that the system can be classified by the relative
homotopy group, and the topological invariant is a half-integer. We term the unexpected
and nontrivial quantum phase as \textquotedblleft quantum anomalous semimetal\textquotedblright .
The topological phase is a synergy of topology of band structure in solid and quantum
anomaly in quantum field theory. The work opens the door towards exploring novel
states of matter with fractional topological charge.

Discovery of quantum Hall effect in 1980s opened an avenue to explore the topological
states of matter and topological materials in condensed matter physics and materials
science \cite{vonKlitzing-80prl}. The topological insulators and superconductors
are classified as a state of matter according to the topology of band structure and
are characterized by the $Z$ and $Z_{2}$ topological invariants \cite{Moore-10nature,hasan2010colloquium,qi2011titsc,armitage2018weyl,shen2012topological,Tokura-18nrp},
which determine the existence of the boundary states around the systems according
to the bulk-boundary correspondence \cite{hasan2010colloquium,qi2011titsc}. Some
topological metals and semimetals were also discovered to possess the Fermi arcs
on the surface of the systems \cite{Murakami-07njp,Wan-11prb,Soluyanov-15nature,Burkov-11prb,Fang-15prb,Bzdusek-16nature,Yang-14nc}.
However, so far all the discovered topological phases of matter and the topological
materials are characterized by a nonzero integer. None of the topological invariant
is a half-integer or fractional \cite{Schnyder-08prb,Chiu-16rmp}. The Nielsen-Ninomiya
or fermion doubling theorem states that a single gapless Dirac fermion cannot be
constructed on a regular lattice in even space-time dimensions with preserving all
of the symmetries: translation invariance, chiral symmetry, locality, and hermiticity
\cite{Nielsen-81plb,Rothe}. By sacrificing one of the presuppositions, unpaired
massless Dirac fermion can be formulated to get rid of the doublers. The chiral anomaly,
one of the fundamental physics in quantum field theory, is closely related to the
theorem. In odd space-time dimensions, the fermion doubling phenomenon in two spatial
dimensions is intimately tied to another quantum anomaly, the so-called parity anomaly,
which can be viewed as the analog to chiral anomaly in even space-time dimensions.
Here we propose a novel type of topological state of matter, termed \textquotedblleft quantum
anomalous semimetal\textquotedblright{} to emphasize its close relation to quantum
anomalies. This phase hosts the Wilson fermions instead of the Dirac fermions. The
gapless Wilson fermions break the chiral or parity symmetry at generic momenta, and
can be realized on lattices. . It is found that the topological phase is classified
by the relative homotopy group and characterized by a half-integer topological invariant.
The half-integer topological invariant leads to a \textcolor{black}{fractional electric
and electromagnetic polarization in one and three dimensions}, and half-quantized
Hall conductance in two and four dimensions with no well-defined boundary states,
forming a novel type of the bulk-boundary correspondence in the topological phase.
An explicit consequence in the one-dimensional (1D) phase is the prediction of the
transfer of one half of elementary charge $e/2$ in a topological charge pumping,
which demonstrates the distinction of the phase to all other existing topological
phases and materials.

\paragraph*{Model and Wilson fermions}

The massless Wilson fermions can be realized as a consequence of fermion regularization
on a lattice, which can be described by the modified Dirac equation on a $d$-dimensional
hyper-cubic lattice \cite{Wilson-75,Rothe,Qi2008topological,shen2012topological},
\begin{equation}
H=\sum_{i=1}^{d}\left(\frac{\hbar v}{a}\sin k_{i}a\alpha_{i}-\frac{4b}{a^{2}}\sin^{2}\frac{k_{i}a}{2}\beta\right)
\end{equation}
with the lattice space $a$ and effective velocity $v$ where $\alpha_{i}$ and $\beta$
are the Dirac matrices. Its conduction band and valence band $E_{\pm}=\pm\sqrt{\sum_{i=0}^{d}f_{i}^{2}(\mathbf{k})}$
with $f_{0}(\mathbf{k})=-\frac{4b}{a^{2}}\sum_{i=1}^{d}\sin^{2}\frac{k_{i}a}{2}$
and $f_{i}(\mathbf{k})=\frac{\hbar v}{a}\sin k_{i}a$ ($i=1,...,d$) touches at $\mathbf{k}=0$
to form a single Dirac cone of massless fermions (See Fig. 1a and b). A continuous
model is valid by taking $\sin k_{i}a\simeq k_{i}a$ and also used in the following.
In the case of $d=1,3$, it is known that the chiral symmetry is broken explicitly,
in which the chirality operator $\mathcal{C}=e^{-i\frac{\pi}{4}(d-1)}\prod_{i=1}^{d}\alpha_{i}$
does not commute with $\beta$. The Hamiltonian also exhibits a global sublattice
symmetry, $\Gamma H\Gamma^{-1}=-H$ with $\Gamma=e^{-i\frac{\pi}{4}(d+1)}\beta\prod_{i=1}^{d}\alpha_{i}$.
In the case of $d=2$,\textcolor{black}{{} the parity operator is defined as $\mathcal{P}=\sigma_{x}M$
and under $M$: $x\rightarrow x$ and $y\rightarrow-y$ }\cite{Deser1982Topologically}.
The $b$ term changes its sign under $\mathcal{P}$ and breaks the parity symmetry
explicitly. Thus the Wilson fermions break the chiral or parity symmetry explicitly
and avoid fermion doubling problem \cite{Rothe}. The symmetry is restored when $k_{i}\rightarrow0$
near the degenerate point. Consequently, the fermion doubling phenomenon is intimately
tied to quantum anomaly, \textit{i.e.} the chiral anomaly for $d=1,3$ \cite{adler1969axial,Bell69nca}
and the parity anomaly for $d=2,4$ \cite{Niemi1983Axial,Haldane1988Model}.

\paragraph*{Classification of relative homotopy group}

In the $d$-dimensional space, the Brillouin zone is topologically equivalent to
a torus: $\mathbf{k}\in T^{d}$. Assume that the band crossing occurs at a single
point in the momentum space $\{\mathbf{w}\}$ which serves as the ``monopole''
of the gauge field. To present a homotopy classification, we need to remove the degenerate
point to avoid the singularity, which will change the topology of the Brillouin zone.
We assume there are $n$ occupied and $n$ unoccupied Bloch bands for each momentum.
On the complement $T^{d}\backslash\{\mathbf{w}\}$, one can define the Q-matrix $Q(\mathbf{k})=2P(\mathbf{k})-I$
in terms of the projection operator $P(\mathbf{k})=\sum_{a=1}^{n}|u_{a}(\mathbf{k})\rangle\langle u_{a}(\mathbf{k})|$,
where $|u_{a}(\mathbf{k})\rangle$ are the occupied Bloch wave-functions. $I$ is
a $2n\times2n$ identity matrix. For a given system, $Q(\mathbf{k})$ defines a continuous
map from the Brillouin zone $T^{d}\backslash\{\mathbf{w}\}$ to a specific topological
manifold $M$, such that the Brillouin zone boundary surrounding the degenerate point
($\cong S^{d-1}$) is mapped to a submanifold $X\subset M$ (see an example in two
dimensions as illustrated in Fig. 1e). Mathematically, the classification of topological
semimetal phases is equivalent to distinguish distinct classes for all such mappings
which is given by the relative homotopy group $\pi_{d}(M,X)$ \cite{Hatcher2002algebraic,Mermin1979topological}.
In the following, we will identify the topological invariants with elements of the
relative homotogy group for quantum anomalous semimetals in even (odd) spatial dimensions
that parity (chiral) symmetry is broken for generic momenta but restored surrounding
the degenerate point.

In even spatial dimensions, we consider the Hamiltonian has no constraint other than
being Hermitian. The Q-matrix thus defines a map from $T^{d}\backslash\{\mathbf{w}\}$
into the the complex Grassmannian $G_{n,2n}(\mathbb{C})=\frac{U(2n)}{U(n)\times U(n)}$.
The topological invariants characterizing distinct topological phases in this symmetry
class are the first Chern number $\nu_{2D}$ and second Chern number $\nu_{4D}$
for two and four dimensions, respectively \cite{Schnyder-08prb,Qi2008topological}.
On the infinitesimal boundary surrounding the degenerate point, the parity symmetry
is restored that the space of $Q$ is restricted to $U(n)$. By using the exact sequence
\cite{Hatcher2002algebraic}, the relative homotopy group can be derived as: $\pi_{d}[G_{n,2n}(\mathbb{C}),U(n)]\simeq\mathbb{Z}\oplus\mathbb{Z}$.
Due to the presence of parity symmetry near the degenerate point, the Brillouin zone
boundary contributes a half-integer to $\nu_{2D}$ and $\nu_{4D}$. Thus the Chern
numbers for the parity anomalous semimetals in two and four dimensions can be expressed
as $\nu_{2D/4D}=N_{1}+\frac{1}{2}N_{2}$ with $N_{1}$ and $N_{2}$ all integers.
In odd spatial dimensions, we restrict our discussions in systems with sublattice
symmetry which indicates that we can find a unitary matrix $\Gamma$ anticommuting
with the Hamiltonian, $\Gamma H\Gamma^{-1}=-H$. As a consequence, the Q-matrix can
be brought into an off-diagonalized form with the off-diagonal component as $q(\mathbf{k})$
which defines a map from the Brillouin zone onto $U(n)$. In this case, a winding
number $w_{1D/3D}$ can be defined to characterize distinct topological classes.
The restored chiral symmetry around the degenerate point further restricts $q(\mathbf{k})\in G_{a,n}(\mathbb{C})$
with $a\le n$ on the boundary of the Brillouin zone. The chiral anomalous semimetal
is then classified by the relative homotopy group $\pi_{d}[U(n),G_{a,n}(\mathbb{C})]\simeq\mathbb{Z}\oplus\mathbb{Z}$.
Due to the chiral symmetry, the contribution of the Brillouin zone boundary to the
topological invariants is half-quantized. As a consequence, the winding number for
the chiral anomalous semimetals in one and three dimensions can be obtained as $w_{1D/3D}=N_{1}+\frac{1}{2}N_{2}$
with $N_{1}$ and $N_{2}$ all integers.

The existing classification theory of semimetals is based on the properties of the
band structure near the crossing points\cite{Chiu-16rmp}. For quantum anomalous
semimetal, there are two topological integers to classify the matter, as given by
the relative homotopy group. One characterizes the topology of the bands on the sphere
$S^{d-1}$ surrounding the crossing point and the other one characterizes the bands
on the high energy scale. The quantum anomalous semimetals here host the massless
Wilson fermions instead of the conventional Dirac fermions. The two types of fermions
are similar near the crossing point but distinctly different at higher energy scales.
The gapless Wilson fermions provide the simplest example of the quantum anomalous
semimetals in various dimensions since the $b$ term vanishes in the vicinity of
the degenerate point. The topological invariant for gapless Wilson fermions is simply
one half, $-\frac{1}{2}\mathrm{\mathrm{sgn}}(b)$, which only depends on the sign
of the symmetry broken coefficient $b$.

\paragraph*{1D solvable model and topological half-charge pumping}

For a periodically driven quantum two-level system, if the energy gap remains open,
the final state evolves back to the initial one during a cyclic adiabatic process,
the accumulated geometric phase is gauge invariant and experimentally measurable
\cite{Berry1984quantal}. Geometric phases are key to understanding numerous physical
effects, such as the electric polarization \cite{Resta1992theory,King-Smith-93prb}
and anyonic fractional statistics \cite{Nayak2008nonabelian}. The topological invariants
can be expressed in terms of these geometric phases which characterize the parallel
transport of the ground state upon cyclic changes of system parameters (time $t$
or wave vector $k$ in the crystal band) \cite{Xiao2010Berry}. The time evolution
of the two level system also reveals the novel topological property of the massless
Wilson fermions. Here, we consider a solvable two-level system,
\begin{equation}
H_{1D}(t)=\frac{\hbar\omega_{0}}{2}\sin\omega t\sigma_{x}+\hbar\omega_{0}\sin^{2}\frac{\omega t}{2}\sigma_{y},
\end{equation}
which is periodic with a time $T(=2\pi/\omega)$, $H_{1D}(t+T)=H_{1D}(t)$. Eq. (2)
is equivalent to the Hamiltonian (1) in 1D if $\omega t$ is replcaed by $ka$. The
time evolution of this system is governed by the Schrodinger equation $i\frac{\partial\Psi(t)}{\partial t}=H(t)\Psi(t)$.
The instantaneous energy eigenvalues are $E_{\chi}=\chi\hbar\omega_{0}\sin\frac{\omega t}{2}$
with $\chi=\pm1$. The two bands cross at time $t=0$ or $T$. The model possesses
the glide reflection symmetry $\mathcal{G}(\omega t)=\left(\begin{array}{cc}
0 & e^{-i\omega t}\\
1 & 0
\end{array}\right)$ such that $\mathcal{G}(\omega t)H_{1D}(t)\mathcal{G}^{-1}(\omega t)=H_{1D}(t)$.
The symmetry generator has the relation $\mathcal{G}^{2}(\omega t)=e^{-i\omega t}$,
and its eigenvalues are $\chi e^{-i\omega t/2}$. Using the eigenstates of $\mathcal{G}(\omega t)$
as the basis, and solving the time-dependent Schrodinger equation, it is found that,
in the adiabatic condition of $\alpha=8\omega_{0}/\omega\rightarrow+\infty$ that
the system varies with time very slowly comparing with the band width $\hbar\omega_{0}$,
the state is always stuck to the eigenstate of $\mathcal{G}(\omega t)$, i.e., the
adiabatic theorem is still valid for this gapless system protected by the glide reflection
symmetry. Consequently, the system evolves back to the initial state and the Berry
phase $\pi$ is gained after two periods of time, $\Psi(t=2T)=e^{i\pi}\Psi(t=0)$
as shown in Fig. 2a. At $t=T$, $\Psi(t=T)=e^{i\frac{\pi}{2}}e^{i\alpha/2}\mathcal{G}(\alpha)\Psi(t=0)$.
The phase $\alpha$ is attributed to the dynamic phase of of the system. If the initial
state is one eigenstate of $\mathcal{G}(0)$, then at $t=T$ it will evolve into
another eigenstate of $\mathcal{G}(0)$. For a large but finite $\alpha$ the transition
probability to the initial state at $t=T$ is found to be $\frac{4}{\alpha\pi}$,
which approaches zero in the adiabatic condition. This reflects the non-Abelian topological
property of the 1D system \cite{Wilzcek-84prl,Zhang-17pra}.

Furthermore, a striking feature of 1D Wilson fermions is a realization of the transfer
of one half of elementary charge $e/2$ in a very slow and periodical modulation
in time. The Thouless charge pumping \cite{Thouless-83,Niu-90prl} was first proposed
for a gapped system, in which the transferred charge is always an integer of elementary
charge in an adiabatic cyclic evolution, and was observed experimentally recent years
\cite{Nakajima-16np,Lohse-16np}. A half-quantized pumping rate in a quantum spin
driven by two-harmonic incommensurate drives was proposed recently \cite{Crowey-20prl}.
Here the quantum anomaly of massless Wilson fermions makes it possible to realize
a half-charge transfer in one periodic modulation in time. Let us consider 1D tight-binding
Hamiltonian in a time dependent modulation as shown in Fig. 2b (upper panel),
\begin{equation}
H(t)=\sum_{n}\left[(v(t)+(v_{0}+v(t))(-1)^{n})c_{n}^{\dagger}c_{n+1}+h.c.+w(t)(-1)^{n}c_{n}^{\dagger}c_{n}\right]\label{eq:model}
\end{equation}
where $v_{0}$ is real constant. The modulating hoping strength $v(t)=v_{0}\sin^{2}\frac{\omega t}{2}$
and potential $w(t)=w_{0}\sin(\omega t)$ form a cyclic evolution in the parameter
space shown in Fig. 2b (lower panel) such that the Hamiltonian is also periodic in
time: $H(t)=H(t+T)$. By transforming into momentum space, Eq. (\ref{eq:model})
becomes $H(k,t)=w_{0}\sin(\omega t)\sigma_{z}+v_{0}\sin(ka)\sigma_{y}+2v_{0}(\sin^{2}\frac{ka}{2}+\sin^{2}\frac{\omega t}{2})\sigma_{x}$.
In two dimensional parameter space $(k,t)$, it is equivalent to Eq. (1) in 2D up
to a basis transformation. The energy evolution of the system in the adiabatic condition
is plotted in Fig. 2c and d for the massless and massive cases. Only the two states
around the zero energy evolve in the period of $2T$, and swap their energy signs
at $t=T$, and all other states evolve in the period of $T$ as expected in the adiabatic
evolution for an isolated system.

The charge that flows through the system during one period is given by $\Delta Q=\int_{-T/2}^{T/2}dt\frac{dP_{x}(t)}{dt}$
where $P_{x}$ is the charge polarization defined as the shift of the electron center-of-mass
position away from the lattice sites and is only well-defined modulo 1. It is evaluated
by the many-body polarization formula using all instantaneous occupied states at
time $t$ \cite{King-Smith-93prb}. The time evolution of the energy spectrum in
Eq. (\ref{eq:model}) with a periodical boundary condition for two period $2T$ is
plotted in Fig. 2c. The band gap is closed at $t^{*}=0$. For $t\ne t^{*}$, the
system is fully gapped and the calculated polarization $P_{x}$ is presented in Fig.
2e. The total pumped charge in a single period $T$ is given by $\Delta Q=-\frac{1}{2}\mathrm{sgn}(w_{0})$,
which equals the winding number around the band crossing point in the (1+1)-dimensional
parameter space. Due to the circumvents of fermion doubling problem, the gapless
time $t^{*}$ that corresponds to the appearance of the degenerate point is no longer
paired, which guarantees the quantization of the total pumped charge $\Delta Q$
as a half-integer instead of an integer in a gapped system. This $2T$ periodicity
of polarization evolution for quantum anomalous semimetal is completely different
from the gapped cases that the energy gap remains open during a cyclic change of
$t$ as shown in Fig. 2d. As illustrated in Fig. 2f, the gapping of the quantum anomalous
semimetal leads to two distinct spectrums of polarization which are indicated by
red (trivial) and blue (nontrivial) circles and both exhibit a $T$ periodicity.

\paragraph*{Generalized bulk-boundary correspondence}

The bulk-boundary correspondence lies at the heart of the topological phases. For
example, in the quantum Hall effect the Chern number in quantum Hall conductivity
means the number of the chiral edge states around the boundary \cite{Hatsugai-93prl}.
The half-quantized Hall conductivity in a parity anomalous semimetal here does not
mean the existence of one half of the edge state, but reveals the existence of chiral
edge current although the energy band gap is closed. To this end, we evaluate the
local density of states (LDOS) at the position near the two edges and the middle
position of a strip of two-dimensional (2D) sample with a sufficient width to avoid
the finite size effect. Along the $x$-direction, we take the periodic condition,
and the wave vector $k_{x}$ is a good quantum number. The LDOS can be evaluated
as a function of $k_{x}$ and the position $y$, $\rho(k_{x},y)$ \cite{Chu-11prb}
as plotted in Figs. 3 a,b and c, which positions are labelled schematically in Fig.
3d. At the middle of the sample, the LDOS $\rho(k_{x},y=0)$ is even about $k_{x}$,
which indicates that there is no pure current in the bulk without an external field.
At the two points close to the edges $y=\pm L_{y}/2$ ($L_{y}$ is the width of the
stripe), the relative LDOS $\rho_{r}(k_{x},\pm L_{y}/2)$ are plotted in Fig. 3b
and 3c, in which the contribution of the bulk part has been deducted. We find that
the nonzero relative LDOS emerges and maximizes at $E=\pm\hbar vk_{x}$ at the two
edge positions $y=\pm L_{y}/2$. This biased distribution in LDOS indicates that
a chiral current lays at the edge of the sample. This current can be illuminated
directly by calculating the many-body local current density $\langle v_{x}(y)\rangle$
of all occupied states along the sample, where $v_{x}(y)=\frac{\partial H(k_{x},y)}{\partial k_{y}}$.
Current density distribution at two different chemical potentials are presented in
Figs. 3e. Consequently a chiral current may circulate without well-defined edge states
along the boundary in the absence of external field. This forms a novel type of the
bulk-edge correspondence in a quantum anomalous semimetal.

Formation of the bound state around the domain wall also provides an alternative
way to demonstrate the bulk-boundary correspondence. We first consider a 1D static
domain wall that the parameter $b$ has a kink along the $x$ direction., i.e. $b(x)=b_{0}\mathrm{sgn}(x)$.
In addition to the extended bulk states, the exact solution demonstrate that there
always exists a bound state of zero energy localized around the domain wall $\psi(x)=\chi_{y}\sqrt{|\frac{\hbar v}{2b_{0}}|}\exp\left(-|\hbar vx/b_{0}|\right)$
with $\chi_{y}$ the eigen spinor of $\sigma_{y}$, $\sigma_{y}\chi_{y}=\mathrm{sgn}(b_{0})\chi_{y}$.
The result is analogue to the early theoretical prediction of domain wall fermions
given by Jackiw and Rebbi in the context of relativistic field theory\cite{Jackiw1976solitons}.
However in the present situation, the bound state trapped around the domain wall
between topologically distinct regions will coexist with the extended states in the
bulk.The solution of the bound state can be generalized to two and three dimensions.
In the 2D domain wall in Fig. 3f, the momentum $\hbar k_{y}$ is still conserved.
One can obtain a solution of the chiral bound states along the domain wall with a
linear dispersion $E=\mathrm{sgn}(b_{0})\hbar vk_{y}$ , which may carry one quantized
conductance of $e^{2}/h$. These states are located near the domain wall, and decay
exponentially away from the domain wall in space. Furthermore, in the three-dimensional
(3D) case there exists a gapless Dirac cone of fermions with linear dispersion located
at the interface. All these bound states coexist with the bulk states as the systems
have no energy gap. The existence of the bound state is closely related to the fact
that the difference of the topological invariants between the two sides of the domain
wall is always equal to $+1$ or $-1$, although the topological invariants are one
half. It can be viewed as an extension of the index theorem for gapped systems \cite{Chiu-16rmp}
to the gapless systems with fractional indices.

\paragraph*{Quantum anomaly in a domain wall}

Quantum anomaly provides a deep insight into the peculiar transport properties of
the topological phase. To illustrate the quantum anomaly of Wilson fermions we continue
to investigate the 2D domain wall in Fig. 3f that the parameter $b$ has a kink along
the $x$ direction and is constant along $y$ direction. The system is coupled to
a weak electromagnetic potential $A_{\mu}$ via the minimal substitution rule (dictated
by the gauge invariance). The Chern-Simons field theory provides a natural theoretical
framework to describe the properties of the topological phases \cite{Fradkin2013field,Bottcher2019survival,Burkov2019Dirac}.
The Chern-Simons term arises in the effective action of the gauge field $A_{\mu}$
from the fermionic fluctuations in Wilson fermion as a result of the violation of
parity symmetry. The effective action far from the location of the wall (lies along
the $y$ direction) is $S_{\mathrm{eff}}^{\mathrm{CS}}=\frac{1}{4\pi}\int d^{3}xn_{c}(b(x_{1}))\epsilon^{\mu\nu\lambda}A_{\mu}\partial_{\nu}A_{\lambda}$
with space-time coordinate $x^{\mu}=(t,x,y)$ and $\epsilon^{\mu\nu\lambda}$ Levi-Civita
symbol that the Greek indices ($\mu,\nu,$etc.) run over all the space-time indices
$(0,1,2)$. For (2+1)D translational invariant system, the coupling constant $n_{c}$
in Chern-Simons action is equal to the topological invariant $\nu_{2D}$. By taking
the functional derivatives of $S_{\mathrm{eff}}^{\mathrm{CS}}$ with respect to $A_{\mu}$,
the current can be obtained as $\langle J_{\mathrm{CS}}^{\mathrm{\mu}}\rangle=\frac{1}{2\pi}n_{c}(b(x_{1}))\epsilon^{\mu\nu\lambda}\partial_{\nu}A_{\lambda}-\frac{1}{4\pi}\delta(x_{1})\epsilon^{1\mu\lambda}A_{\lambda}$.
However, this action is not invariant under the gauge transformation $A_{\mu}\to A_{\mu}+\partial_{\mu}\Phi(x)$
at the domain wall, because the domain-wall bound states are chiral which also has
consistent chiral anomaly \cite{Callan1984Anomalies,XGWen1991gapless}. The boundary
action associated with the boundary excitations has to be included such that the
total effective action is gauge invariant. Now we consider the electric field parallel
to the domain wall as $E_{1}=0,$ and $E_{2}=E$. Then the Hall current of the anomalous
Hall state $J_{1}=-\frac{e^{2}}{h}n_{c}(b(x_{1}))E$ flows towards the domain wall
and the longitudinal current along the domain wall is $J_{2}=\frac{e^{2}}{h}E$.
Here we do not take into account the longitudinal minimal conductivity from the 2D
massless fermions. The conductance along the domain wall is quantized. Here the bulk
Hall coefficient $n_{c}=-\frac{1}{2}\mathrm{sgn}(b(x_{1}))$ is half-quantized and
the Hall currents from both sides flow toward or outward the domain wall. The quantized
charge current along the wall can be understood as a consequence of the convergence
of two anomalous Hall currents because of the conservation of the total charge current.
The boundary states suffer from the chiral anomaly as $\partial_{\mu}\langle J^{\mu}\rangle=-\frac{1}{2\pi}\epsilon^{\rho\lambda}\partial_{\rho}A_{\lambda}$
where the Greek indices only run over the space-time indices $(0,2)$, \textit{i.e.}
the charge conservation is broken at the domain wall since the current can leak into
the bulk through the bulk quantum Hall effect. Thus the half-quantization of the
bulk Hall conductance at the two sides of the domain wall is manifested as the quantization
of the chiral anomaly coefficient along the domain wall, which is consistent with
the solution of the chiral states in the previous section. This effect also reveals
the bulk-edge correspondence in this topological phase.

\paragraph*{Possible realization}

The quantum anomalous semimetals can be realized in two alternative ways, one is
the accidental band crossing and another is the band crossing protected by additional
crystalline symmetries. At the critical transition point between the conventional
and topological insulators, the accidental band crossing may give rise to the Wilson
fermions by fine tuning the band gap, for example, the HgTe quantum well grown at
a critical thickness where the band gap vanishes \cite{Buttner-11np} and the strain-controlled
narrow gapped $\mathrm{ZrTe}_{5}$ \cite{Mutch-19SciAdv}. The topological phases
may be stable against sufficient weak but short range electronic interactions and
random mass at least for $d=2$ and 3 based on the scaling and renormalization group
analysis.

Here we take a quasi-1D system as an example to discuss the stability of the band
crossing and the evasion of fermion doubling problem by additional crystalline symmetry.
Generally, the symmetry-enforced band crossings can be movable at some high symmetry
line or pinned at a particular high-symmetry point \cite{Bzdusek-16nature,Elcoro2020arxiv}.
In 1D, the movable band crossing can be protected by the nonsymmorphic crystal symmetry
which is a combination of a point-group symmetry with a translation of a fractional
Bravais lattice. For a biparticle system with glide mirror symmetry $\bar{M}_{y}$,
the square of $\bar{M}_{y}$ is a lattice translation of one unit cell. The bipartite
lattice further possesses the sublattice symmetry when only the nearest neighbor
hopping are included. Consequently, the energy eigenstates are actually the instantaneous
eigenstates of $\bar{M}_{y}$. Upon shifting the momentum by one reciprocal lattice
vector, the eigenvalues of $\bar{M}_{y}$ remains unchanged, but the two energy branches
must cross odd times in the Brillouin zone. One needs to go through the Brillouin
zone twice to get back to the same eigenvalue. Therefore a \textit{global} topological
invariant $w_{1D}$ can be introduced to characterizes the symmetry-enforced band
crossing which measures the winding of the eigenvalue of $\bar{M}_{y}$ as going
through the Brillouin zone. The quantum anomalous semimetal in 2D can be realized
in semi-magnetic topological slab \cite{Mogi2021arxiv}. Consider a time reversal
invariant topological insulator slab, if the surface states are gapped by magnetic
doping or proximity effect at one surface while the surface states at the opposite
surface remain gapless, an unpaired gapless Dirac cone can be realized in this quasi-2D
system. Finding a potential candidate in 3D remains a major challenge and we leave
it for future research. A promising alternative way to realize the topological phase
is in designed artificial systems, such as cold atoms, photonic/acoustic metamaterials,
and circuit networks, which provide good platforms to simulate various topological
phases in solid state physics.

\paragraph*{Acknowledgements}

This work was supported by the Research Grants Council, University Grants Committee,
Hong Kong under Grant No. 17301220, and the National Key R\&D Program of China under
Grant No. 2019YFA0308603.

\paragraph*{Author Contribution}

All authors contributed to performing the calculations and the analysis of the results.
SQS was responsible for the supervision of the project. SQS and BF wrote the manuscript
with suggestions from all authors.

\paragraph*{Competing interests}

The authors declare no competing interests.

\paragraph*{Materials \& Correspondence}

Extended data is available for this paper at {[}XXXX{]}. Correspondence and requests
for materials should be addressed to SQS.

\pagebreak{}

\begin{figure}
\centering{}\includegraphics[width=12cm]{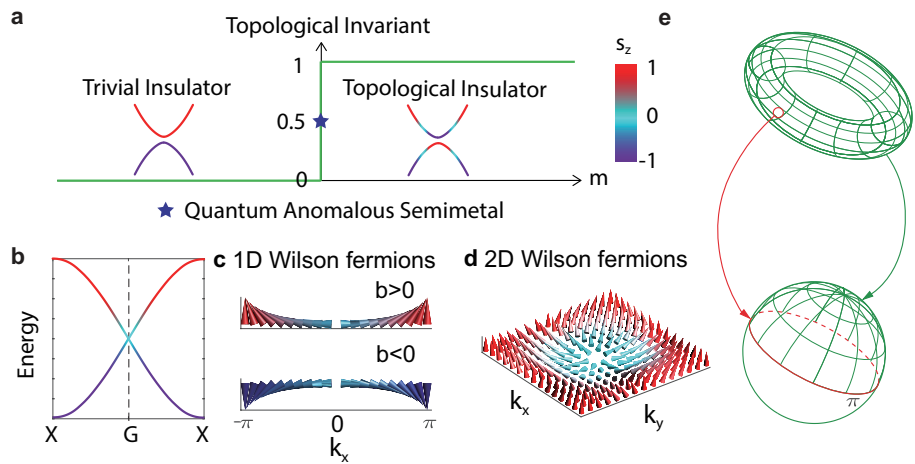} \caption{\textbf{Quantum anomalous semimetal.} a. The phase diagram and topological invariant
of trivial insulator, quantum anomalous semimetal and topological insulator. b. The
dispersion of the Wilson fermions along the wave vector $k_{x}$ and keeping all
other $k_{i}=0$. c. The spin texture of 1D Wilson fermions. d. The spin texture
of 2D Wilson fermions. e. The relative homotopy group mapping of band structure and
spin texture of the parity anomalous semimetal. The whole torus on the first Brillouin
zone maps to the semi-Bloch sphere (green line) and the small loop around crossing
point to the equator of the Bloch sphere (red line).}
\end{figure}

\begin{figure}
\includegraphics[width=12cm]{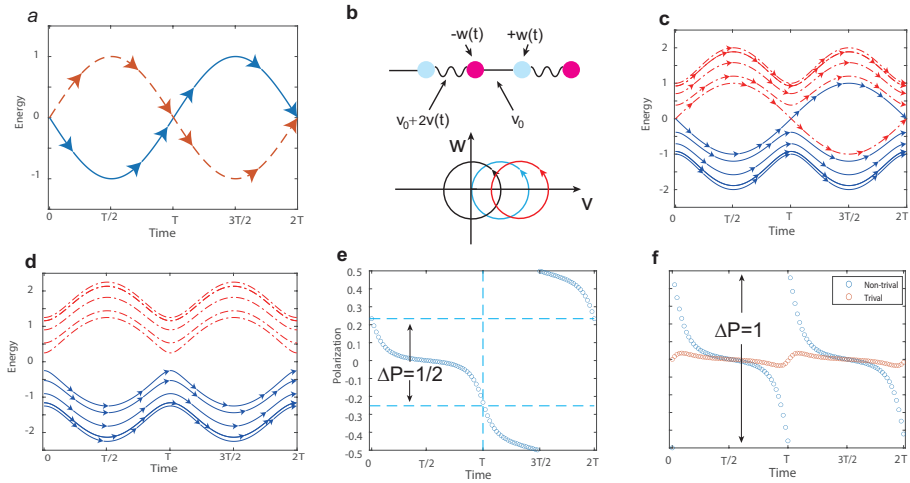}

\caption{\textbf{Topological half-charge Pumping}. a. The two-period time evolution in the
two-level system. b. The sketch of a half-charge pumping chain with $v(t)=v_{0}\sin^{2}\frac{\omega t}{2}$
and $w(t)=w_{0}\sin\frac{2\pi t}{T}$ and the cyclic evolution in $v$-$w$ space.
The black and red circles mean for the topologically nontrivial case $v(t)=-\frac{v_{0}}{2}+v_{0}\sin^{2}\frac{\omega t}{2}$
and trivial case $v(t)=m+v_{0}\sin^{2}\frac{\omega t}{2}$ with $m>0$. c. The energy
dispersion for the gapless one-dimensional chain in Eq. \ref{eq:model} in two time
period $T$. d. The energy dispersion for the gaped one-dimensional chain in Eq.
\ref{eq:model} with a nonzero mass in $v(t)=m+v_{0}\sin^{2}\frac{\omega t}{2}$
in two time period $T$. e. Electric polarization as a function of $t$ in two time
period $T$ for the gapless one dimensional chain. f. Electric polarization as a
function of $t$ in one time period $T$ for the gaped cases of $v(t)=m+v_{0}\sin^{2}\frac{\omega t}{2}$
and $v(t)=-\frac{v_{0}}{2}+v_{0}\sin^{2}\frac{\omega t}{2}$.}
\end{figure}
\begin{figure}
\includegraphics[width=12cm]{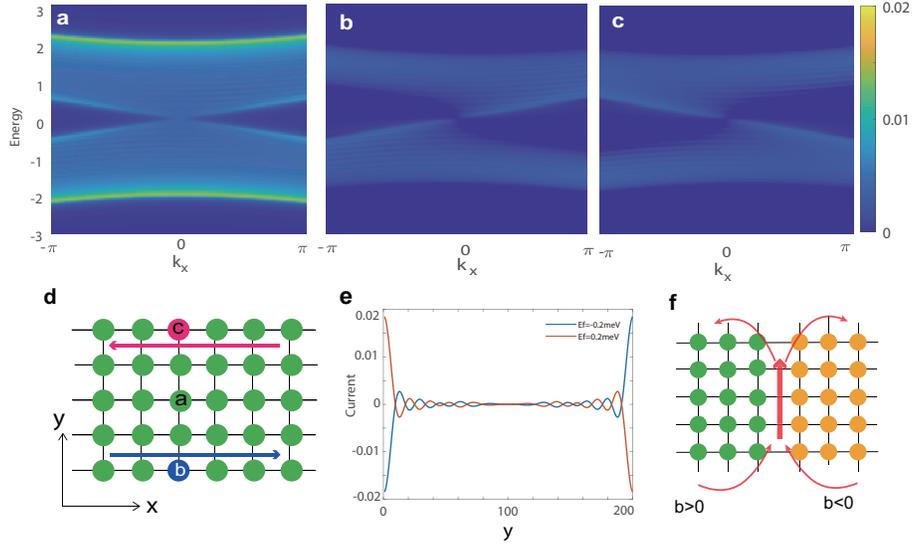}\caption{\textbf{The local density of states and edge current distribution in the absence
of external field}. a. The local density of states at the middle of the sample. b.
The relative local density of states at the top edge (orange dot) $\rho_{r}(k_{x},y=L_{y})$.
c. The relative local density of states at the bottom edge (blue dot) $\rho_{r}(k_{x},y=1)$.The
relative local density of states means that the contribution from bulk is already
deducted, $\rho_{r}(k_{x},1)=\rho(k_{x},1)-\rho(k_{x},0)$ and $\rho_{r}(k_{x},L_{y})=\rho(k_{x},L_{y})-\rho(k_{x},0)$.
d. A schematic of a stripe sample with the labelled positions for a, b, and c. e.
The current density distribution for two different Fermi levels slight deviating
from the half filling. f. The Hall and longitudinal currents along the domain wall
in two dimensions. Here $L_{y}=200$, $v=1$ and $b_{0}=0.5$.}
\end{figure}

\end{document}